\documentclass[sigconf]{acmart}

\AtBeginDocument{%
  }

\usepackage{array}
\usepackage[table]{xcolor}  
\usepackage{subcaption}     
\usepackage{caption}        
\usepackage{enumitem}
\usepackage{tikz}
\usepackage{varwidth}
\usepackage{xcolor}
\usepackage{graphicx}
\usepackage{amsthm}
\usepackage{siunitx} 
\usepackage{csquotes}       
\usepackage{pgfplotstable}
\usepackage{pgfplots}        
\pgfplotsset{compat=1.17}    
\usetikzlibrary{shadows.blur,shapes.geometric,arrows.meta,patterns}

\definecolor{bloom1}{RGB}{66, 133, 244}   
\definecolor{bloom2}{RGB}{52, 168, 83}    
\definecolor{bloom3}{RGB}{251, 188, 5}    
\definecolor{bloom4}{RGB}{234, 67, 53}    
\definecolor{bloom5}{RGB}{147, 51, 234}   
\definecolor{bloom6}{RGB}{20, 184, 166}   

\definecolor{viridis1}{RGB}{68,1,84}      
\definecolor{viridis2}{RGB}{59,82,139}    
\definecolor{viridis3}{RGB}{33,145,140}   
\definecolor{viridis4}{RGB}{94,201,97}    
\definecolor{viridis5}{RGB}{253,231,37}   
\definecolor{viridis6}{RGB}{255,127,0}    
\definecolor{viridis7}{RGB}{220,20,60}    

\newcommand{\hlNonMale}[1]{{\setlength{\fboxsep}{1.5pt}\colorbox{viridis1!20}{#1}}}
\newcommand{\hlFirstGen}[1]{{\setlength{\fboxsep}{1.5pt}\colorbox{viridis2!20}{#1}}}
\newcommand{\hlLowIncome}[1]{{\setlength{\fboxsep}{1.5pt}\colorbox{viridis3!20}{#1}}}
\newcommand{\hlESL}[1]{{\setlength{\fboxsep}{1.5pt}\colorbox{viridis4!20}{#1}}}
\newcommand{\hlPriorExp}[1]{{\setlength{\fboxsep}{1.5pt}\colorbox{viridis5!20}{#1}}}
\newcommand{\hlCSCE}[1]{{\setlength{\fboxsep}{1.5pt}\colorbox{viridis6!20}{#1}}}
\newcommand{\hlBLNPI}[1]{{\setlength{\fboxsep}{1.5pt}\colorbox{viridis7!20}{#1}}}

\newcommand{\leveltag}[3]{%
  \tikz[baseline=(tag.base)]{
    \node[
      rounded corners=3pt,
      fill=gray!10,
      draw=gray!50,
      line width=0.3pt,
      inner xsep=4pt,
      inner ysep=1pt,
      font=\bfseries
    ] (tag) {
      #1: #2
    };
  }%
}

\newcommand{\participantid}[2]{%
  \normalfont\small\itshape --- #1\,|\,\textsc{#2}%
}

\newcommand{\inlinequote}[2]{%
  \vspace{0.2em}%
  \begin{quote}
  \smash{\raisebox{-3.5ex}{\fontsize{28}{28}\selectfont\textcolor{gray!60}{``}}}\normalsize\itshape\space #1\upshape
  \parseParticipant{#2}
  \end{quote}
  \vspace{0.2em}%
}

\newcommand{\parseParticipant}[1]{%
  \def\temp{#1}%
  \expandafter\parseParticipantHelper\temp\relax
}
\def\parseParticipantHelper P #1, #2\relax{%
  \participantid{P #1}{#2}%
}

\author{Annapurna Vadaparty}
\orcid{0009-0002-4370-764X}
\affiliation{%
  \institution{University of California - San Diego}
  \city{La Jolla}
  \state{California}
  \country{USA}}
\email{avadaparty@ucsd.edu}

\author{David H. Smith IV}
\orcid{0000-0002-6572-4347}
\affiliation{%
  \institution{Virginia Tech}
  \city{Blacksburg}
  \state{Virginia}
  \country{USA}}
\email{dhsmith4@vt.edu}

\author{Samvrit Srinath}
\orcid{0009-0009-4135-9849}
\affiliation{%
  \institution{University of California - San Diego}
  \city{La Jolla}
  \state{California}
  \country{USA}}
\email{sasrinath@ucsd.edu}

\author{Mounika Padala}
\orcid{0009-0006-9943-0159}
\affiliation{%
  \institution{University of California - San Diego}
  \city{La Jolla}
  \state{California}
  \country{USA}}
\email{mpadala@ucsd.edu}

\author{Christine Alvarado}
\orcid{0000-0003-1182-1069}
\affiliation{%
  \institution{University of California - San Diego}
  \city{La Jolla}
  \state{California}
  \country{USA}}
\email{cjalvarado@ucsd.edu}

\author{Jamie Gorson Benario}
\orcid{0000-0002-0385-4357}
\affiliation{%
  \institution{Google}
  \city{Chicago}
  \state{Illinois}
  \country{USA}}
\email{jamben@google.com}

\author{Daniel Zingaro}
\affiliation{%
 \institution{University of Toronto Mississauga}
 \city{Toronto}
 \state{Ontario}
 \country{Canada}}
 \email{daniel.zingaro@utoronto.ca}

\author{Leo Porter}
\affiliation{%
 \institution{University of California - San Diego}
 \city{La Jolla}
 \state{California}
 \country{USA}}
\email{leporter@ucsd.edu}

\begin{abstract}
Generative AI (GenAI) models have broad implications for education in general, impacting the foundations of what we teach and how we assess. This is especially true in computing, where LLMs tuned for coding have demonstrated shockingly good performance on the types of assignments historically used in introductory CS (CS1) courses. As a result, CS1 courses will need to change what skills are taught and how they are assessed. Computing education researchers have begun to study student use of LLMs, but there remains much to be understood about the ways that these tools affect student outcomes.
In this paper, we present the design and evaluation of a new CS1 course at a large research-intensive university that integrates the use of LLMs as a learning tool for students. We describe the design principles used to create our new CS1-LLM course, our new course objectives, and evaluation of student outcomes and perceptions throughout the course as measured by assessment scores and surveys. Our findings suggest that 1) student exam performance outcomes, including differences among demographic groups, are largely similar to historical outcomes for courses without integration of LLM tools, 2) large, open-ended projects may be particularly valuable in an LLM context, and 3)
students predominantly found the LLM tools helpful, although some had concerns regarding over-reliance on the tools. 
\end{abstract}

\begin{document}

\title[Integrating LLMs and Evaluating Student Outcomes in Intro CS]{Integrating Large Language Models and Evaluating Student Outcomes in an Introductory Computer Science Course}

\maketitle

\fancyhead{}  



\section{Introduction}\label{sec1}

If the advent of the internet beckoned in the information age, the advent of widely available generative AI (GenAI) models represents our entering the age of AI. The widespread availability of these models has led to concerns over academic integrity~\cite{elkins2020can,grassini2023shaping,rahman2023chatgpt,lo2023impact}. In computing education, these concerns are made more acute by the widespread availability of GenAI-powered programming assistants. 

The role of GenAI in computing courses is under intense study~\cite{prather2025hype}, but there has been little research on the outcomes of fully redesigned courses integrating GenAIs. Current studies have focused on small-scale interventions, student perceptions, and behavior when using GenAI tools~\cite{kazemitabaar2024codeaid, 10.1145/3639474.3640076}. While these studies provide preliminary insights, the impact of GenAI integration in large-scale university courses remains unclear.

We have designed a new introductory CS (CS1) course that fully incorporates student use of GenAI tools. In this paper, we investigate how student outcomes in this course compare to historical norms, both with respect to international benchmarks~\cite{simon2016benchmarking} and across demographic groups that have been of historical interest to computing education~\cite{hagan2000does, guo2018non, thompson2022high, neda2024investigating, salguero2021understanding}. Similarly, to guide future iterations of courses that may seek to adopt our approach, we evaluate student perceptions of the course and the GenAI tools they employed. To these ends, we address the following research questions:
\begin{enumerate}
    \item[\textbf{RQ1:}] How did students' performance when learning to program with GenAI compare to historical, pre-GenAI benchmarks?
    \item[\textbf{RQ2:}] What were students' perceptions of learning to program with GenAI tools?
    \item[\textbf{RQ3:}] How do students' performance and perceptions vary across different student populations in a course that centers GenAI?  
\end{enumerate}

\section{Literature Review}\label{sec2}

\subsection{Generative AI in CS1}\label{subsec:llms_in_cs1} 

The advent of widely available Generative AI (GenAI) models is reshaping introductory computer science education. If a GenAI model were a student, it would rank among the top performers in CS1 assessments~\cite{finnie2022robots}. Furthermore, the widespread adoption of tools like GitHub Copilot in industry, used by over 20 million developers~\footnote{https://techcrunch.com/2025/07/30/github-copilot-crosses-20-million-all-time-users/}, signals that students entering the workforce will need proficiency in these tools~\footnote{https://github.blog/2023-10-10-research-quantifying-github-copilots-impact-on-code-quality/}. As a result, traditional learning objectives are being reexamined, as discussed by \citet{becker2023programming} in ``\textit{Programming is Hard -- Or at Least it Used to Be}:'' 

\begin{quote}
  \textit{``AI-generated code presents both opportunities and challenges for students and educators in introductory programming and related courses. The sudden viability and ease of access to these tools suggest educators may be caught unaware or unprepared for the significant impact on education practice resulting from AI-generated code. We therefore urgently need to review our educational practices in the light of these new technologies.''}
\end{quote}
They, and others, suggest that the rise of GenAI will give way to novel pedagogical approaches that reduce barriers to programming and allow students to focus on higher-level concepts such as problem solving, algorithmic design, and developing larger applications earlier in their education~\cite{feng2025automation, porter2024learn}. 
Though it would be inappropriate to consider these tools as a panacea for all the barriers students face when learning to program, these models, and systems built on top of them, have the opportunity to offer additional support to students through AI tutoring systems~\cite{liffiton2023codehelp, kazemitabaar2024codeaid}, 
auto-grading systems~\cite{smith2024evaluating, alkafaween2025automating}, 
and content generation~\cite{ling2024automatic, cambaz2024use}. 

Prior studies have explored students' use of GenAI tools in computing, such as through prompt-writing tasks that require understanding a programming problem to prompt functional code~\cite{smith2024prompting, denny2024prompt, denny2024explaining}. 
Lab-based (quasi-experimental) studies have observed student behaviors and surveyed attitudes toward using GenAI in programming tasks~\cite{kazemitabaar2024codeaid, 10.1145/3639474.3640076}, 
showing generally positive attitudes and similar performance compared to peers without GenAI tools. However, these findings were generally in lab settings, and it remains unclear how integrating GenAI into CS1 course instruction affects student learning. This study aims to address that gap.

\subsection{Differences in Student Outcomes in CS1}\label{subsec:student_outcomes} 

CS1 courses often do not equally support all students, with women, Black, Latine/x/a/o, and Native American students being underrepresented~\cite{camp2017generation}. Demographic factors and pre-college educational access can influence CS1 outcomes, making it crucial to address structures that may hinder students from diverse backgrounds. This section reviews literature highlighting these disparities to ensure that new courses, like our GenAI integrated CS1 course, are designed to reduce or not worsen these gaps.

\subsubsection{Prior Experience}

Although CS1 is intended for beginners, it often favors students with prior programming experience, who consistently outperform their peers~\cite{hagan2000does}. This dynamic can discourage less experienced students~\cite{petersen2016revisiting}, and we do not yet know how LLMs might affect this pattern.

\subsubsection{English as a Second Language}

English proficiency influences CS1 success, as programming instruction and code are primarily in English~\cite{guo2018non}. Non-native speakers often face language-based challenges that can impact performance~\cite{10.1145/1113847.1113888, ben2018correlation}. As a community, we need to understand whether and how such challenges present when working with LLMs.

\subsubsection{Socioeconomic Status}

Students from higher SES backgrounds tend to perform better in CS1~\cite{thompson2022high}, partly due to related factors like prior experience and parental education~\cite{neda2024investigating, sirin2005socioeconomic}. These advantages convey access to resources and academic norms that can widen equity gaps. Curriculum design must consider and address these disparities.

\subsubsection{Race and Gender}

Racial disparities in CS1 outcomes are linked to differences in pre-college access and systemic biases in instruction~\cite{margolis2008stuck, katz2003gender}. While there are inconsistent findings related to gender-based performance gaps~\cite{taylor1994exploration, beyer2014women}, factors like self-efficacy and belonging strongly influence outcomes~\cite{lishinski2021all, krause2021relationship}. We examine how integrating LLMs might influence these dynamics.

\begin{figure*}
  \centering
  \includegraphics[width=0.95\textwidth]{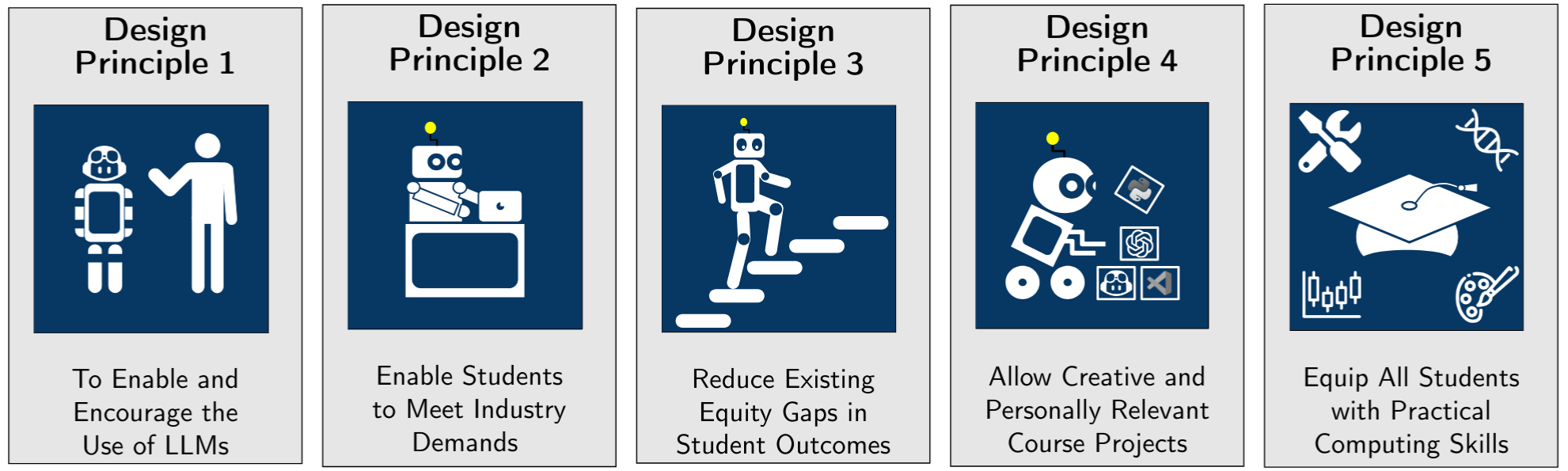}
  \vspace{-0.25cm}
  \caption{Design Principles for the course }
  \label{fig:design_principles}
  \vspace{-0.05cm}
\end{figure*}

\subsection{Changing CS1 Learning Objectives in Light of GenAI}

With LLMs increasingly capable of generating code for complex problems, instructors must reconsider which skills remain critical and which new ones should be taught in introductory programming. Prompting is one such emerging skill, requiring students to articulate a program’s purpose in natural language. This natural language reasoning parallels “Explain in Plain English” (EiPE) questions, where students describe code behavior in words; studies suggest that EiPE questions can be used to develop code comprehension and prompting abilities in tandem\cite{smith2024evaluating, denny2024explaining}. 

While EiPE has long been included in traditional (non-LLM) CS1 contexts, other skills—particularly debugging and testing—have taken on new importance in the LLM era. Despite LLMs’ capabilities, \citet{dakhel2023github} found that Copilot can still produce buggy code, emphasizing the importance of students' mastery of programming fundamentals to evaluate LLM output. Identifying such bugs relies on testing practices, which some instructors are bringing to their CS1 courses ~\cite{fernandez2024cs1}. Interviews with instructors~\cite{lau2023ban} further indicate this shift toward new skills in courses that teach programming with GenAI. This study's course design addresses these needs.

\section{Course Design}\label{sec5}

In this section, we outline the design of our new CS1 course. We include its guiding principles and their justification(Section~\ref{subsec:design_principles}), the learning objectives derived from these principles (Section~\ref{subsec:llm_learning_objectives}), and our course implementation choices (Section~\ref{subsec:course_design}). 

\subsection{Design Principles for CS1-LLM}\label{subsec:design_principles}

We adopted the design principles for our CS1 course from \citeauthor{vadaparty-CS1LLM}~\cite{vadaparty-CS1LLM}. Here we briefly summarize these design principles (Figure~\ref{fig:design_principles}).
Importantly, they are \textit{complementary} to existing best practices in CS1. 

\paragraph{Design Principle 1 -- Enable and Encourage Use of LLMs:} 

Throughout the course, instruction and assessments were designed to include LLM-integrated activities  emphasizing skills like open-ended design, prompting, reading, and evaluating LLM-generated code~\cite{prather2023robots}. While students in LLM-integrated environments will be able to generate code without strong programming fundamentals, they will need the aforementioned skills in order to work with this code that they have generated. 

\paragraph{Design Principle 2 -- Enable Students to Meet Industry Demands:}

There is often a gap between what traditional computer science coursework prepares students for and the skills required in industry~\cite{craig2018listening}. While debate continues about how much CS1 should emphasize industry preparation, students must still master tools and concepts needed for their majors and careers. Our approach emphasizes code reading, problem decomposition, and testing—skills critical as students work with and assess LLM-generated code that may initially contain errors~\cite{dakhel2023github}. These skills are reinforced through both homework and non-LLM exam questions to ensure that students are building these skills independently, and through project-based work that encourages the use of LLMs. 

\paragraph{Design Principle 3 -- Reduce Equity Gaps in Student Outcomes:}

Making CS1 courses more equitable across underrepresented groups has been a central topic of research within computer science education. Despite best practices that help address performance gaps between under- and highly-represented groups ~\cite{crouch2001peer, werner2004pair}, these gaps persist~\cite{salguero2020longitudinal,salguero2021understanding}. Student-to-student interaction  practices such as pair programming and peer instruction have also been shown to make CS1 courses more equitable~\cite{crouch2001peer, werner2004pair}. Our CS1 course---both before and after the incorporation of GenAI---employs a number of these techniques including peer instruction and contextualized computing. It additionally includes open-ended projects that promote student agency and choice; we suspected that GenAI would be particularly beneficial on such large tasks.

\paragraph{Design Principle 4 -- Allow Creative and Personally Relevant Course Projects:}

Building on \textit{Design Principles 2 and 3}, we designed projects that reflect real-world, project-based work and allow students to explore creative, personally meaningful topics. Many real-world contexts involve larger, less-structured tasks than the Many Small Programs (MSPs) approach of traditional CS1 homework questions~\cite{allen2018weekly}. We use the One Large Program (OLP) model~\cite{allen2021many}, enabling deeper integration of skills and authentic applications. LLMs help reduce the barrier to completing OLPs, allowing students to engage with realistic problem decomposition and pursue personalized goals.

\paragraph{Design Principle 5 -- Equip All Students With Practical Computing Skills:}

As computing spreads across disciplines, accessible introductory courses are essential. Many address this through domain-specific applications such as ``Media Computation"~\cite{guzdial2003media}, where students manipulate images and sounds. Beyond CS1, CS+X programs help students apply computing within other fields~\cite{sloan2020cs}. LLMs offer a new way to teach practical computing, helping students tackle advanced, relevant tasks early without traditional programming mastery.
\vspace{-0.4em} 
\subsection{Learning Objectives}\label{subsec:llm_learning_objectives}

Even before the advent of GenAI models led to tectonic shifts 
in computing education, there was little consensus on the central learning objectives of an introductory computing course~\cite{becker2019cs1}. The appropriateness of objectives depends on the student population for which those outcomes are calibrated. 

We adapted our course to include specific learning objectives related to the use of LLMs in programming and to de-emphasize traditional learning objectives focused on writing code from scratch.  Shown below are those objectives organized using Bloom's taxonomy~\cite{krathwohl2002revision}:

\vspace{8pt}
\noindent\leveltag{Level 1}{Knowledge}{bloom1} Students will be able to define: nondeterminism, Large Language Model (LLM), prompt, prompt engineering, code correctness, problem decomposition, and top-down design.

\vspace{6pt}
\noindent\leveltag{Level 2}{Comprehension}{bloom2} Students will illustrate the workflow used when programming with an AI assistant and describe the purpose of common Python programming features, including: 1) Variables, 2) Conditionals, 3) Loops, 4) Functions, 5) Lists, 6) Dictionaries, and 7) Modules.

\vspace{6pt}
\noindent\leveltag{Level 3}{Application}{bloom3} Students will apply prompt engineering to influence code generated by an AI assistant.

\vspace{6pt}
\noindent\leveltag{Level 4}{Analysis}{bloom4} Students will analyze and trace a Python program to determine or explain its behavior, divide a programming problem into subproblems as part of top-down design, and debug a Python program to locate bugs.

\vspace{6pt}
\noindent\leveltag{Level 5}{Synthesis}{bloom5} Students will design open- and closed-box tests to determine whether code is correct, identify and fix bugs in Python code, modify Python code to perform a different task, and write complete and correct Python programs using top-down design, prompting, testing, and debugging.

\vspace{6pt}
\noindent\leveltag{Level 6}{Evaluation}{bloom6} Students will judge whether a program is correct using evidence from testing and debugging.

\subsection{Course Implementation}\label{subsec:course_design}

\begin{table}[t]
  \centering
  \caption{Course Schedule}\label{tab:course_schedule}
  \vspace{-0.35cm}
  \definecolor{headercolor}{RGB}{70,130,180}
\definecolor{rowcolor1}{RGB}{245,245,245}
\definecolor{rowcolor2}{RGB}{255,255,255}
\definecolor{bordercolor}{RGB}{0,0,0}
\setlength{\arrayrulewidth}{0.8pt}
\arrayrulecolor{bordercolor}
\begin{tabular}{cl}
\hline
\rowcolor{headercolor}
\textcolor{white}{\textbf{Week}} & \textcolor{white}{\textbf{Topic(s)}} \\
\hline
\rowcolor{rowcolor1} 1 & Functions and Working with Copilot\\
\rowcolor{rowcolor2} 2 & Variables, Conditionals, Memory Models\\
\rowcolor{rowcolor1} 3 & Loops, Strings, Testing, VSCode Debugger\\
\rowcolor{rowcolor2} 4 & Loops, Lists, Files, Problem Decomposition \\
\rowcolor{rowcolor1} 5 & Intro to Data Science, Dictionaries\\
\rowcolor{rowcolor2} 6 & Revisit Problem Decomposition and Testing \\
\rowcolor{rowcolor1} 7 & Intro to Images, PIL, Image Filters\\
\rowcolor{rowcolor2} 8 & Copying Images, Intro to Games and Randomness\\
\rowcolor{rowcolor1} 9 & Large Game Example\\
\rowcolor{rowcolor2} 10 & Python Modules and Automating Tedious Tasks\\
\hline
\end{tabular}

\end{table}

\begin{table}[t]
  \centering
  \caption{Course Assessments and Weights}\label{tab:assignment_weights}
  \vspace{-0.35cm}
  \definecolor{headercolor}{RGB}{70,130,180}
\definecolor{rowcolor1}{RGB}{245,245,245}
\definecolor{rowcolor2}{RGB}{255,255,255}
\definecolor{bordercolor}{RGB}{0,0,0}
\setlength{\arrayrulewidth}{0.8pt}
\arrayrulecolor{bordercolor}
\begin{tabular}{lcc}
\hline
\rowcolor{headercolor}
\textcolor{white}{\textbf{Assessment Type}} & \textcolor{white}{\textbf{Component}} & \textcolor{white}{\textbf{Weight}} \\
\hline
\rowcolor{rowcolor1} Formative (35\%) & Peer Instruction Participation & 5\% \\
\rowcolor{rowcolor2} & Reading Quizzes & 5\% \\
\rowcolor{rowcolor1} & Homework & 15\% \\
\rowcolor{rowcolor2} & Labs & 10\% \\
\hline
\rowcolor{rowcolor1} Summative (65\%) & Projects & 10\% \\
\rowcolor{rowcolor2} & Mid Quarter Exams & 30\% \\
\rowcolor{rowcolor1} & Final Exam & 25\% \\
\hline
\end{tabular}

\end{table}

Our new CS1 course takes place at a large research-focused university in the United States. The course schedule and weighting of assignments appear in Tables~\ref{tab:course_schedule} and ~\ref{tab:assignment_weights}. The LLM environment introduced was GitHub Copilot in Microsoft's Visual Studio Code, selected because of their seamless integration and accessibility via the GitHub Student Developer Pack (free of charge for students). Below we expand on the course components. 

\paragraph{Textbook:} 
Our teaching staff used the textbook ``Learn AI-Assisted Python Programming with GitHub Copilot and ChatGPT", assigning weekly readings and graded quizzes. The book combines traditional CS1 topics with LLM-related material like effective Copilot usage and prompt engineering~\cite{porter2024learn}.

\paragraph{Lectures:}
Lectures included mini-lectures, Live Coding~\cite{selvaraj2021live},
 and Peer Instruction~\cite{porter2016multi}, covering typical CS1 topics along with discussion of Copilot interactions. Instructors highlighted incorrect Copilot responses, prompting students to identify issues, and demonstrated Copilot Chat in VS Code for answering code-related questions.

\paragraph{Formative Assessments:}
Our course used PrairieLearn for homework and assessments, leveraging features like problem variants and autograding~\cite{west2015prairielearn}. Homework included multiple choice, short answer, Explain in Plain English~\cite{murphy2012explain}, Parsons Problems~\cite{parsons2006parson}, debugging, and code writing. Students were encouraged to work without Copilot for most problems (in preparation for summative assessments) but could use it if needed. Some coding questions were intended for LLM usage, featuring a Copilot-enabled workspace.

\paragraph{Summative Assessments:}
The course included four 50-minute mid-quarter exams featuring code tracing, explaining~\cite{venables2009closer}, Parsons Problems~\cite{parsons2006parson}, and small coding tasks. Three exams excluded Copilot, but one question on the second exam allowed Copilot access. We originally intended to include Copilot sections on all exams following exam 1, but technical issues in the login process limited inclusion.

The final exam consisted of three parts, all of which were proctored.
\begin{enumerate}
    \item \textbf{Multiple-choice} (90~min, 70\%) --- tracing, explaining, testing, debugging, Parsons, and short-answer items.
    \item \textbf{Non-Copilot coding} (45~min, 15\%) --- four coding questions in PrairieLearn on lab computers.
    \item \textbf{Copilot-assisted coding} (45~min, 15\%) --- one large problem on personal or lab computers, with sample test cases and partial credit. Variants involved either analyzing a new dataset or creating a restricted spell-check tool.
\end{enumerate}

\section{Analysis and Results}\label{sec6}

Our analysis includes comparing CS1-LLM student performance to international CS1 benchmarks (Section~\ref{sec:perf_comp}), examining students' perceptions of programming with LLMs (Section~\ref{sec:perceptions}), and analyzing performance differences across demographics (Section~\ref{sec:perf_CS1LLM_demographics}).

\subsection{RQ1) CS1-LLM vs CS1 Grades}\label{sec:perf_comp}

In this section, we compare CS1-LLM student performance to international CS1 course averages reported in~\citet{simon2016benchmarking} to assess our students' understanding of programming fundamentals while learning with LLMs. 

\subsubsection{Data Collection and Analysis}\label{sec:perf_data_collection_analysis}

\begin{table*}[t]
    \centering
    \caption{Questions used from \citet{simon2016benchmarking}}\label{tab:simon_questions}
    \vspace{-0.25cm}
    
    \definecolor{headercolor}{RGB}{70,130,180}
    \definecolor{footercolor}{RGB}{230,230,230}
    \definecolor{rowcolor1}{RGB}{245,245,245}
    \definecolor{rowcolor2}{RGB}{255,255,255}
    \definecolor{bordercolor}{RGB}{0,0,0}
    
    \footnotesize
    \resizebox{\textwidth}{!}{%
    \setlength{\arrayrulewidth}{0.8pt}
    \arrayrulecolor{bordercolor}
    \begin{tabular}{llp{7cm}ll}
    \hline
    \rowcolor{headercolor}
    \textcolor{white}{\textbf{Q\#}} & 
    \textcolor{white}{\textbf{Topic}} & 
    \textcolor{white}{\textbf{Task}} & 
    \textcolor{white}{\textbf{Question Type}} & 
    \textcolor{white}{\textbf{Skill Required}} \\ \hline
    \rowcolor{rowcolor1} Q1 & Boolean Expressions & Trace a boolean expression with relational and logical operators. & Multiple Choice & Trace code \\
    \rowcolor{rowcolor2} Q2 & Assignment and Sequence & Trace simple assignment and sequence. & Multiple Choice & Trace code \\
    \rowcolor{rowcolor1} Q3 & Conditional & Trace a nested if statement. & Short Response & Trace code \\
    \rowcolor{rowcolor2} Q4 & Conditional & Trace a nested if/elif to find a variable’s value. & Multiple Choice & Trace code \\
    \rowcolor{rowcolor1} Q5 & Iteration and Arrays & Trace a while loop with conditionals and arrays. & Short Response & Trace code \\
    \rowcolor{rowcolor2} Q6 & Iteration and Arrays & Determine the outcome of a for loop over an array. & Multiple Choice & Explain code \\
    \rowcolor{rowcolor1} Q7 & Iteration & Describe the purpose of loop-based code. & Short Response & Explain code \\ \hline
  \end{tabular}%

    }
    \vspace{-0.25cm}
\end{table*}

On the final exam, we included questions from the \citet{simon2016benchmarking} study (Table~\ref{tab:simon_questions}). With the exception of Q7, a short answer ``Explain in Plain English'' question, all questions were multiple-choice or fill in the blank. We graded Q7 out of 4 points. We awarded full points for correct relational responses, 3 points for correct but less abstract responses, 2 points for responses that explained the code line by line, and 1 or 0 points for serious misunderstandings.

In total, we had 535 responses to all questions except Q7 ($n=250$) for which we had to remove the first cohort of test-taker responses due to an error in the question. We compare the performance of students in the CS1-LLM course to that of the international averages presented in the \citet{simon2016benchmarking} study.

\subsubsection{Results}\label{sec:perf_results}

%
%
\begin{figure}[t]
    \centering
    \includegraphics[width=\columnwidth]{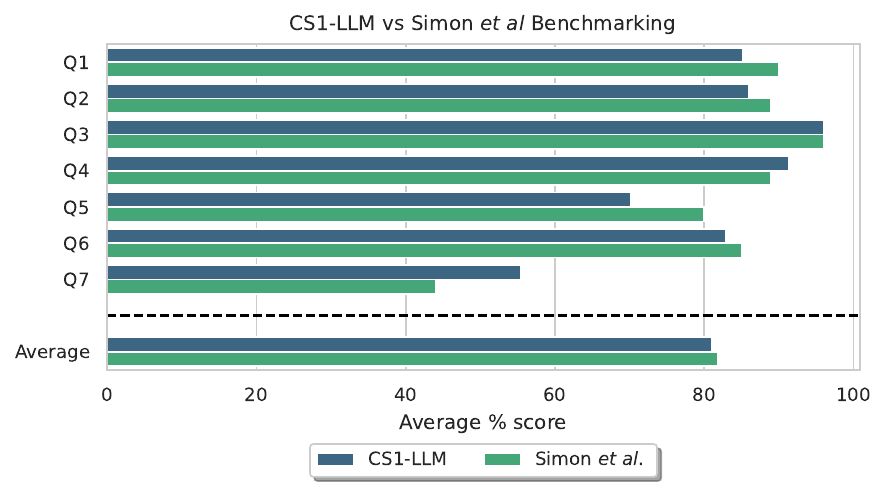}
    \vspace{-0.75cm}
    \caption{\citet{simon2016benchmarking} Benchmarking Data Results}
    \label{fig:simon_benchmarking}
\end{figure}

Overall, the results of the question level comparisons suggest that our
students learning with GenAI was largely comparable to past students
learning without GenAI. Students in our GenAI-centered course
performed similarly to those from historical international averages
(Figure~\ref{fig:simon_benchmarking}). The one question type in which students
in the CS1-LLM course performed notably better (11\%) than the international
average was on the short answer ``Explain in Plain English'' question (Q7).
This may suggest that students in CS1-LLM were better equipped to understand
and articulate the purpose of code, perhaps as a result of practicing how to prompt LLMs and evaluate their output.

\subsection{\textbf{RQ2) Students' Perceptions of CS1-LLM}}\label{sec:perceptions}

In this section we explore students perceptions of learning to code with Copilot.

\subsubsection{Data Collection and Analysis}

To gain an understanding of student perceptions of the course, we distributed
two surveys, one midway through the quarter (\textit{mid}) and the other at the
end of the quarter prior to the final exam (\textit{end}).  The \textit{mid}
and \textit{end} surveys asked students a variety of Likert questions along
with the following open-ended question:
\begin{quote}
``\textit{How did you feel about working with Copilot as you learned to program this quarter?}''
\end{quote}
Two researchers qualitatively analyzed responses to this question from both the
mid- (400 respondents) and the end-of quarter surveys (315 respondents). To
build a code book, they independently inductively coded 
the mid-survey responses in batches of
100 and used negotiated agreement to resolve disagreements. After reaching saturation with no new codes emerging, they then independently coded an additional 100 responses to establish
inter-rater reliability ($\alpha=0.86$) using Krippendorff's
Alpha~\cite{krippendorff2011computing}. The remaining mid- and
end-of-quarter survey responses were then deductively coded using the code
book. We present the themes that emerged from this process and the codes that
underlie those themes.

\subsubsection{Results}

We highlight four themes that emerged from the qualitative analysis: Theme 1)
\textit{Copilot is helpful for supporting novices to complete programming
tasks}, Theme 2) \textit{Copilot can be helpful for learning}, Theme 3)
\textit{Some difficulties emerge when working with Copilot}, and Theme 4)
\textit{Students can feel overreliant on Copilot}. 

\paragraph{\textbf{Theme 1. Copilot is helpful for supporting novices to complete programming tasks.}} 
Students were largely positive regarding programming with Copilot, with 69.2\% of mid-quarter survey respondents and 42.2\%
of end-quarter survey respondents ($N_{mid}=272$, $N_{end}=133$) mentioning that
Copilot was helpful. Several students indicated that it
\textit{saved time} ($N_{mid}=21$ [5.3\%], $N_{end}=6$ [1.9\%]), and \textit{helped
them fix errors} ($N_{mid}=15$ [3.8\%],$N_{end}=5$ [1.6\%]): 

\begin{itemize}[leftmargin=1em, rightmargin=1em, label={}, itemsep=0.1em, parsep=0.4em]

    
    \item \inlinequote{I feel that \textbf{Copilot is very useful in terms of expediting tasks when
    coding}, and also \textbf{playing the role of an aid when not knowing what
    to do}.}{P 79, Mid}
    
    \item \inlinequote{I feel like copilot was a great tool that was able to assist me.
    \textbf{Whenever I'm developing code and feel like I ran into a bug, I use
    co pilot to help debug} and \textbf{it saves me so much time} on my projects
    and assignments.}{P 531, Mid}

    
\end{itemize}

One way that students specified that Copilot affords coding support was through
\textit{help with questions/hints} about code ($N_{mid}=32$ [8.1\%], $N_{end}=18$ [5.7\%]). Students reported finding
it helpful to ask Copilot questions regarding the functionality and syntax of code. Additionally, some students found Copilot helpful
for \textit{overcoming obstacles} such as not knowing where to get started,
overcoming some form of ``coding writer's block", or more generally, when they
were simply stuck while coding ($N_{mid}=37$ [9.4\%], $N_{end}=19$ [6\%]):

\begin{itemize}[leftmargin=1em, rightmargin=1em, label={}, itemsep=0.1em, parsep=0.4em]

  \item \inlinequote{Copilot was extremely helpful and \textbf{when I was stuck} it would greatly help when it came to \textbf{explaining code}.}{P 415, End}
\end{itemize}

In particular, several students indicated that they found Copilot particularly
\textit{helpful as individuals with no prior programming experience}, as it
helped them navigate coding for the first time  ($N_{mid}=12$ [3.1\%],$N_{end}=4$ [1.3\%]).

\begin{itemize}[leftmargin=1em, rightmargin=1em, label={}, itemsep=0.1em, parsep=0.4em]
  \item \inlinequote{Programming is a completely new concept to me so I feel as though \textbf{Copilot was a good helping hand to help get my toes in the water} and experience the coding world for a first timer.}{P 429, Mid}

\end{itemize}

Students found Copilot helpful for their programming experience in a variety of
ways including ideating about code, understanding it, and fixing it. These features may have been especially helpful in assisting novice programmers
to get started with coding. This suggests that one of Copilot's key benefits, from students' perspectives, is its ability to reduce friction in expressing their ideas in code. 

\paragraph{\textbf{Theme 2. Copilot can be helpful for learning. }}

In reflecting on the connection between their learning process and the use of
Copilot, many students ($N_{mid}=45$ [11.5\%], $N_{end}=26$ [8.3\%])  reported
that Copilot was \textit{effective for learning and/or understanding} code or
course material: 

\begin{itemize}[leftmargin=1em, rightmargin=1em, label={}, itemsep=0.1em, parsep=0.4em]

    \item \inlinequote{It \textbf{helped to understand more complicated problems} that I would not have been able to do myself. After reading through the generated code,\textbf{ I get a better understanding of how to solve the problem}.}{P 366, End} 
    
    \item \inlinequote{I liked Copilot because it was \textbf{handy when I needed help understanding how to do something} or when I was lost I would then look at the code produced and work backwards.}{P 415, Mid}
    
\end{itemize}

Students found Copilot's outputs to be helpful in understanding how to approach problems, and in understanding how code functions. 
Given these encouraging findings, it may be worthwhile to explore the particular code-related topics that students found Copilot useful for, the specifics of how they used it, and how these findings can translate to curricular or tool design.

A small subset of students specified that they found unfamiliar code that Copilot generated to be helpful by allowing them to \textit{learn from new techniques} ($N_{mid}=19$ [4.8\%], $N_{end}=6$ [1.9\%]). In particular, students stated that they were able to learn functions and syntax with which they were previously unfamiliar.

\begin{itemize}[leftmargin=1em, rightmargin=1em, label={}, itemsep=0.1em, parsep=0.4em]
    \item \inlinequote{Working with Copilot has been great for helping me understand how to format my code, adjust errors, and \textbf{learn about new coding techniques}.}{P 479, End}
    \item \inlinequote{I feel as though it was an amazing tool that helped me in situations that I felt stuck in. It also helped me to \textbf{diversify my coding syntax} which helped the coding process go by more smoothly.}{P 5, Mid}
\end{itemize}

Our findings suggest that some students find these new, unfamiliar code constructs helpful, and instructors may consider ways to intentionally introduce students to these more advanced constructs that LLMs often generate. 

\paragraph{\textbf{Theme 3. Some difficulties emerge when working with Copilot. }} 

While several students found it helpful to learn from new code in Copilot's
outputs as described in Theme 2, some did find its unfamiliar outputs to be
confusing. 
A small number of students found it \textit{difficult to interpret
output} ($N_{mid}=6$ [1.5\%], $N_{end}=4$ [1.3\%]), which sometimes led to
difficulties when the output was incorrect and \textit{needed to be
corrected} ($N_{mid}=7$ [1.8\%], $N_{end}=10$ [3.2\%]): 

\begin{itemize}[leftmargin=1em, rightmargin=1em, label={}, itemsep=0.1em, parsep=0.4em]
    \item \inlinequote{I feel like Copilot has been helpful, however, because of my lack in programming knowledge, I sometimes \textbf{struggle with understanding the complex code that Copilot had generated}.}{P 90, Mid}
    \item \inlinequote{It does help giving me a start but \textbf{I need to modify it a lot sometimes} to get to what I want.}{P 449, End}
\end{itemize}

As LLMs may provide code that uses advanced features of the language or code that has bugs, it is not surprising that some students reported being challenged by encountering such code.  We recommend educators warn students that this can happen and offer suggestions on how to respond when it does.  For example, educators can encourage students to use Copilot to explain the unfamiliar code.

\paragraph{\textbf{Theme 4. Students can feel overreliant on Copilot.}} 
Although most students felt that Copilot was helpful for learning as described
in Theme 2, many students expressed concerns about being \textit{overreliant on
Copilot} for solving coding problems ($N_{mid}=94$ [23.9\%], $N_{end}=80$
[25.4\%]), stating that they used Copilot more than they felt
that they should while coding. Many also mentioned that they felt that it was
\textit{hindering their learning} of the course content and fundamentals of
programming ($N_{mid}=59$ [15\%], $N_{end}=44$ [14\%]):  

\begin{itemize}[leftmargin=1em, rightmargin=1em, label={}, itemsep=0.1em, parsep=0.4em]
    \item \inlinequote{By using Copilot to program, I think that \textbf{it interfered with my ability to code independently because Copilot did all the work for me.} Thus, I struggle to code without Copilot.}{P 29, End}
    \item \inlinequote{Copilot definitely helped me get through assignments, but I feel like I \textbf{relied on it too much} and now \textbf{lack a fundamental understanding of programming basics}. If I were asked to code without Copilot, I wouldn't feel very confident in myself despite doing well in the course.}{P 44, End}
\end{itemize}

Students' feelings surrounding overreliance were evident in several 
comments that \textit{Copilot should be introduced later} ($N_{mid}$ = 15
[3.8\%], $N_{end}$ = 8 [2.5\%]): 

\begin{itemize}[leftmargin=1em, rightmargin=1em, label={}, itemsep=0.1em, parsep=0.4em]
    \item \inlinequote{I feel like there are too many assignments that \textbf{require copilot too early on that I become dependent on copilot}. I think the \textbf{first few weeks should be learning to program without copilot} and later for projects we should use copilot in a limited capacity to help us program complicated functions. I feel like I just ask copilot and it gives me the answer but I don't program a lot without it which is a problem.}{P 282, Mid}
\end{itemize}

While some students felt that the course should introduce Copilot later, some
addressed their dependence on Copilot by \textit{self-regulating their Copilot
use}, often through decreasing the amount of time that they spent using
Copilot. Reports of these self-regulation behaviors emerged in our
end-of-quarter survey ($N_{end}$ = 13 [4.1\%]):

\begin{itemize}[leftmargin=1em, rightmargin=1em, label={}, itemsep=0.1em, parsep=0.4em]
  \item \inlinequote{\textbf{I turned it off after week 2/3}, it should be introduced once the first project rolls around, I know people who can't code because of co-pilot.}{P 478, End} 

    \item \inlinequote{At first, using Copilot interfered with me learning how to program but as \textbf{I began to use it in more moderation} such as attempting to problem solve and work out coding on my own first and \textbf{resorting to Copilot only when I am really really stuck} on a problem has helped.}{P 153, End} 

\end{itemize}

This raises questions of what ``over-reliance'' on Copilot means to students.
Familiarity with GenAI tools was a core learning objective of the course, and
so we wonder why students felt negatively about their reliance on Copilot. It
would be strange, for instance, if students stated that they felt over-reliant
on a debugger after having been introduced to and encouraged to use a debugging
tool. We posit two possible reasons for students feeling as if they were
overreliant on Copilot: 1) lack of access during summative assessments and 2)
beliefs about use of Copilot by professional programmers:

\paragraph{Lack of Access During Summative Assessments:} One factor that likely affected how reliant students felt they should be on Copilot was the instructor decision to restrict access to Copilot during the majority of their summative assessments.  The lack of access on exams may have signaled to students a value in completing certain tasks without the use of Copilot, potentially causing students to see Copilot as a crutch more than a resource.  This is reflected from the several students who expressed \textit{summative preparation concerns} ($N_{mid}=16$ [4.1\%], $N_{end}=6$ [1.9\%]): 

\begin{itemize}[leftmargin=1em, rightmargin=1em, label={}, itemsep=0.1em, parsep=0.4em]
    \item \inlinequote{Although it was helpful, it was really difficult for me on exams, the whole course is based on using AI tools such as copilot for help, yet \textbf{taking it away on exams when we had it in every other type of work seemed a little unfair}.}{P 176, End}

    \item \inlinequote{Its decent, and helpful- but this class takes it too far. We \textbf{need to be able to code without it, OR we can't be quizzed without it}. OR make assignments that are specifically easier and designed to be done without Copilot.}{P 484, End}
\end{itemize}

These comments highlight the difference in access between formative
and summative assessments. As set forth in the course design, the teaching
staff aimed to ensure that students were able to meet core competencies both
with and without access to generative AI. However, limiting access to Copilot on summative exams may
have impacted student perceptions of the role of GenAI, perhaps causing student
overreliance beliefs.

\paragraph{Beliefs about Use of Copilot by Professional Programmers:}
\begin{figure}[tb]
    \centering
    \definecolor{viridis1}{RGB}{68,1,84}       
    \definecolor{viridis2}{RGB}{59,82,139}     
    \definecolor{viridis3}{RGB}{33,145,140}    
    \definecolor{viridis4}{RGB}{94,201,98}     
    
    \begin{tikzpicture}
    \pgfplotstableread{
Row  Rarely  Sometimes  Routinely  Everyday
1    10      33.8       40.3       15.9
    }\frequency
    
    \begin{axis}[
        scale only axis,
        name=ax1,
        xbar stacked,
        xmin=-100,
        xmax=0,
        xtick={-100,-80,-60,-40,-20,0},
        xticklabels={100\%,80\%,60\%,40\%,20\%,0\%},
        xticklabel style={font=\footnotesize},
        ytick=data,
        yticklabels={},
        bar width=0.3cm,
        enlarge y limits={abs=0.1cm},
        width=0.425\columnwidth,
        height=0.8cm,
        axis x line*=bottom,
        axis y line=none,
        grid=major,
        grid style={gray!20}
    ]
        \addplot[fill=viridis2, draw=none] table [x expr=-\thisrow{Sometimes}, y expr=\coordindex] {\frequency};
        \addplot[fill=viridis1, draw=none] table [x expr=-\thisrow{Rarely}, y expr=\coordindex] {\frequency};
    \end{axis}
    
    \begin{axis}[
        scale only axis,
        at=(ax1.south east),
        anchor=south west,
        xbar stacked,
        xmin=0,
        xmax=100,
        xtick={0,20,40,60,80,100},
        xticklabels={0\%,20\%,40\%,60\%,80\%,100\%},
        xticklabel style={font=\footnotesize},
        ytick=data,
        yticklabels={},
        bar width=0.3cm,
        enlarge y limits={abs=0.1cm},
        width=0.425\columnwidth,
        height=0.8cm,
        axis x line*=bottom,
        axis y line=none,
        grid=major,
        grid style={gray!20}
    ]
        \addplot[fill=viridis3, draw=none] table [x=Routinely, y expr=\coordindex] {\frequency};
        \addplot[fill=viridis4, draw=none] table [x=Everyday, y expr=\coordindex] {\frequency};
    \end{axis}
    
    \draw[gray!50, line width=1pt] (ax1.south east) -- (ax1.north east);
    
    \node[above=0.25cm] at (current bounding box.north) {\small\textbf{\textit{If I had to guess, professional programmers use Copilot or similar tools...}}};
    
    \node[below=0cm] at (current bounding box.south) {
        \begin{tikzpicture}[baseline]
        \draw[fill=viridis1] (0,0) rectangle (0.3,0.15);
        \node[right] at (0.32,0.075) {\footnotesize Rarely};
        \draw[fill=viridis2] (1.5,0) rectangle (1.8,0.15);
        \node[right] at (1.82,0.075) {\footnotesize Sometimes};
        \draw[fill=viridis3] (3.3,0) rectangle (3.6,0.15);
        \node[right] at (3.62,0.075) {\footnotesize Routinely};
        \draw[fill=viridis4] (5,0) rectangle (5.3,0.15);
        \node[right] at (5.32,0.075) {\footnotesize Everyday};
        \end{tikzpicture}
    };
    \end{tikzpicture}
    \vspace{-0.75cm}
    \caption{Students perceptions of how often professionals use GenAI tools}\label{fig:prof-genai}
    \vspace{-0.5cm}
\end{figure}
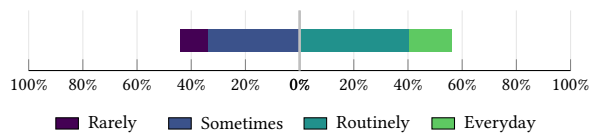 
Another factor that could contribute to feelings of overreliance is that
students may not see Copilot as an industry-standard tool, but instead as a
tool that enables them to complete assignments with or without what they
consider to be a strong understanding of course material. We found that 43.8\%
of students thought that professional programmers only use Copilot or similar
tools rarely or sometimes, while the other 56.2\% thought that professionals
use these tools routinely or every day (Figure~\ref{fig:prof-genai}). It is
possible that with almost half of the class thinking that professional
programmers do not routinely use Copilot, frequent use of--or reliance on--such
tools was not seen as acceptable. These students may consider Copilot to be a
crutch that professional programmers do not or should not need, rather than a
legitimate part of current and future software engineering practices. 

\begin{figure*}[htbp]
  \centering
  \begin{subfigure}[b]{\columnwidth}
    \centering
    \includegraphics[width=\textwidth]{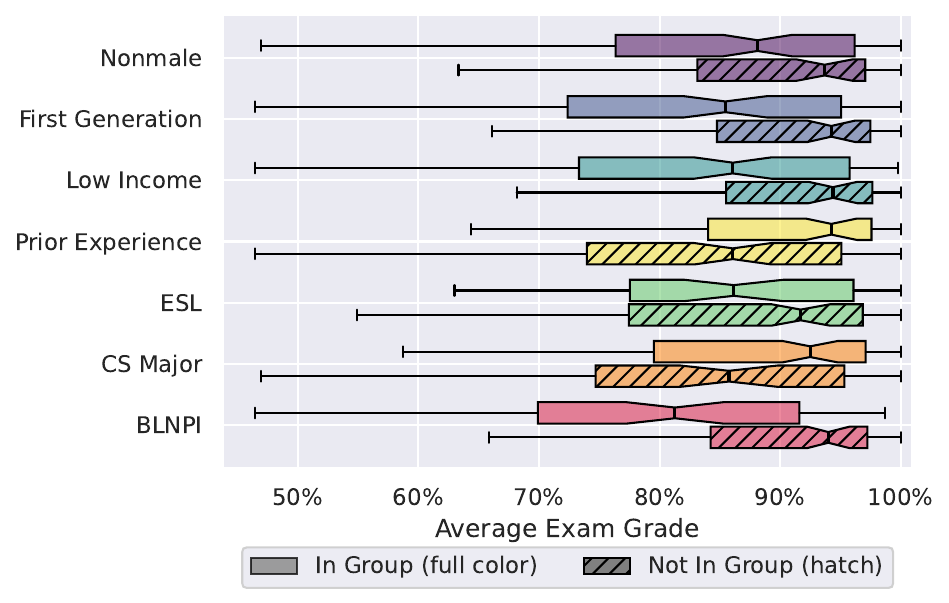} 
  \end{subfigure}
  \hfill
  \begin{subfigure}[b]{\columnwidth}
    \centering
    \includegraphics[width=\textwidth]{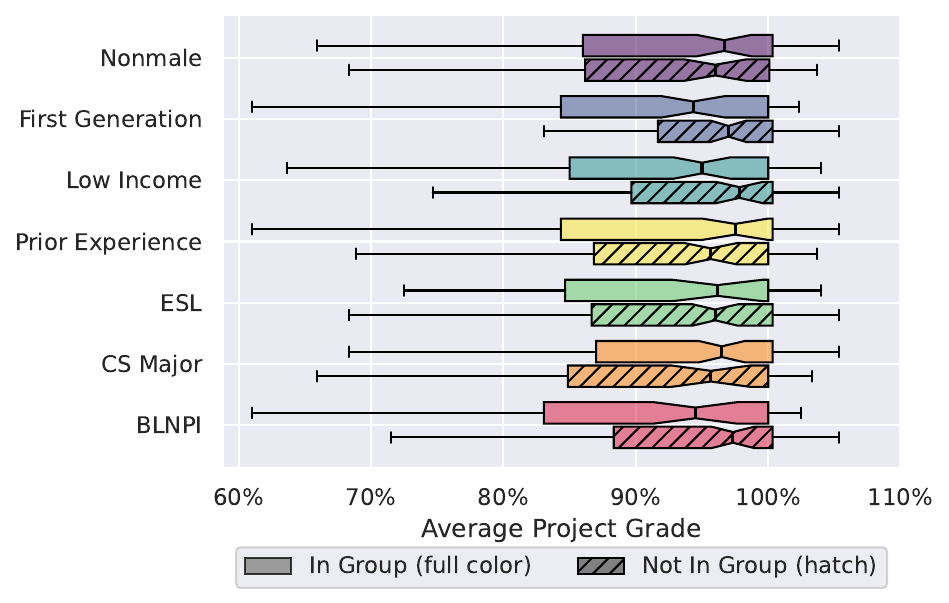} 
  \end{subfigure}
  \vspace{-0.25cm}
  \caption{Average exam and project grades}\label{fig:exam_proj}
\end{figure*}
\begin{table*}[t]
  \footnotesize
  \centering
  \caption{OLS regression results for average exam grades and average project grades by demographic group}\label{tab:grade_proj_ols}
  \vspace{-0.25cm}

  \definecolor{headercolor}{RGB}{70,130,180}
  \definecolor{footercolor}{RGB}{230,230,230}
  \definecolor{rowcolor1}{RGB}{245,245,245}
  \definecolor{rowcolor2}{RGB}{255,255,255}
  \definecolor{bordercolor}{RGB}{0,0,0}

  \begin{subtable}[t]{0.48\textwidth}
    \centering
    \caption{Average Exam Grades}
    \resizebox{\columnwidth}{!}{%
    \setlength{\arrayrulewidth}{0.8pt}
    \arrayrulecolor{bordercolor}
    \begin{tabular}{p{2.85cm} S[table-format=-2.4] S[table-format=1.3] S[table-format=-2.3] S[table-format=1.3, table-space-text-post={***}] c}
      \hline
      \rowcolor{headercolor}
      \textcolor{white}{\textbf{Variable}} &
      {\textcolor{white}{\textbf{$\beta$}}} &
      {\textcolor{white}{\textbf{SE}}} &
      {\textcolor{white}{\textbf{t}}} &
      {\textcolor{white}{\textbf{p}}} &
      \textcolor{white}{\textbf{95\% CI}} \\ \hline
      \rowcolor{rowcolor1} Intercept      & 88.8785 & 2.262 & 39.297 & 0.000\phantom{***} & [84.418, 93.339] \\
      \rowcolor{rowcolor2} \raisebox{0.3em}{\colorbox{viridis1!50}{}} Non-Male (N=119) & -1.9343 & 1.644 & -1.177 & 0.241\phantom{***} & [-5.176, 1.307] \\
      \rowcolor{rowcolor1} \raisebox{0.3em}{\colorbox{viridis2!50}{}} First Gen (N=105)   & -2.3014 & 1.993 & -1.155 & 0.249\phantom{***} & [-6.231, 1.628] \\
      \rowcolor{rowcolor2} \raisebox{0.3em}{\colorbox{viridis3!50}{}} Low Income (N=119)         & -1.8873 & 1.919 & -0.984 & 0.327\phantom{***} & [-5.671, 1.896] \\
      \rowcolor{rowcolor1} \raisebox{0.3em}{\colorbox{viridis5!50}{}} Prior Experience (N=100) &  3.1900 & 1.610 &  1.981 & 0.049*\phantom{**} & [ 0.014, 6.366] \\
      \rowcolor{rowcolor2} \raisebox{0.3em}{\colorbox{viridis4!50}{}} ESL (N=50)        & -0.4636 & 1.835 & -0.253 & 0.801\phantom{***} & [-4.082, 3.154] \\
      
      \rowcolor{rowcolor1} \raisebox{0.3em}{\colorbox{viridis6!50}{}} Computing Major (N=144)      &  3.3696 & 1.780 &  1.893 & 0.060\phantom{***} & [-0.141, 6.880] \\
      \rowcolor{rowcolor2} \raisebox{0.3em}{\colorbox{viridis7!50}{}} BLNPI (N=72)      & -8.4021 & 1.847 & -4.550 & 0.000*** & [-12.044, -4.760] \\ \hline
      \rowcolor{footercolor}
      \multicolumn{6}{c}{$N = 207$ \quad Adj. R$^2$ = 0.216 \quad F = 9.107} \\ \hline
    \end{tabular}%
    } 
  \end{subtable}
  \hfill
  \begin{subtable}[t]{0.48\textwidth}
    \centering
    \caption{Average Project Grades}
    \resizebox{\columnwidth}{!}{%
    \setlength{\arrayrulewidth}{0.8pt}
    \arrayrulecolor{bordercolor}
    \begin{tabular}{p{2.85cm} S[table-format=-2.4] S[table-format=1.3] S[table-format=-2.3] S[table-format=1.3, table-space-text-post={***}] c}
      \hline
      \rowcolor{headercolor}
      \textcolor{white}{\textbf{Variable}} &
      {\textcolor{white}{\textbf{$\beta$}}} &
      {\textcolor{white}{\textbf{SE}}} &
      {\textcolor{white}{\textbf{t}}} &
      {\textcolor{white}{\textbf{p}}} &
      \textcolor{white}{\textbf{95\% CI}} \\ \hline
      \rowcolor{rowcolor1} Intercept      & 92.3976 & 3.328 & 27.762 & 0.000\phantom{***} & [85.835, 98.961] \\
      \rowcolor{rowcolor2} \raisebox{0.3em}{\colorbox{viridis1!50}{}} Non-Male (N=119)     & -0.5270 & 2.419 & -0.218 & 0.828\phantom{***} & [-5.297, 4.243] \\
      \rowcolor{rowcolor1} \raisebox{0.3em}{\colorbox{viridis2!50}{}} First Gen (N=105)    & -3.6446 & 2.932 & -1.243 & 0.215\phantom{***} & [-9.427, 2.137] \\
      \rowcolor{rowcolor2} \raisebox{0.3em}{\colorbox{viridis3!50}{}} Low Income (N=119)          &  1.5766 & 2.823 &  0.558 & 0.577\phantom{***} & [-3.991, 7.144] \\
      \rowcolor{rowcolor1} \raisebox{0.3em}{\colorbox{viridis5!50}{}} Prior Experience (N=100) &  0.5965 & 2.370 &  0.252 & 0.802\phantom{***} & [-4.077, 5.270] \\      
      \rowcolor{rowcolor2} \raisebox{0.3em}{\colorbox{viridis4!50}{}} ESL (N=50)         & -1.1723 & 2.700 & -0.434 & 0.665\phantom{***} & [-6.496, 4.152] \\

      \rowcolor{rowcolor1} \raisebox{0.3em}{\colorbox{viridis6!50}{}} Computing Major (N=114)      &  0.1832 & 2.620 &  0.070 & 0.944\phantom{***} & [-4.983, 5.349] \\
      \rowcolor{rowcolor2} \raisebox{0.3em}{\colorbox{viridis7!50}{}} BLNPI (N=72)       & -2.4990 & 2.717 & -0.920 & 0.359\phantom{***} & [-7.858, 2.860] \\ \hline
      \rowcolor{footercolor}
      \multicolumn{6}{c}{$N = 207$ \quad Adj. R$^2$ = -0.011 \quad F = 0.670} \\ \hline
    \end{tabular}%
    }
  \end{subtable}\label{tab:scores}
  \begin{minipage}{0.95\linewidth}
    \footnotesize \textit{Significance levels:} *** p < 0.001, ** p < 0.01, * p < 0.05.
  \end{minipage}
\end{table*}
\subsubsection{Mixed Perceptions about Copilot: Concurrent Themes}

In open ended responses, a subset of the class reported a tension between the helpfulness of Copilot as a tool for learning and its potential to hinder learning by always providing an immediate answer when they encountered difficulties ($N_{end}$ = 67 [21.3\%]): 

\begin{itemize}[leftmargin=1em, rightmargin=1em, label={}, itemsep=0.1em, parsep=0.4em]
    \item \inlinequote{I feel a bit conflicted, since it has \textbf{helped me write code and
    understand functions at times}, but I also feel that I \textbf{didn't apply
    myself very well} to the course and used Copilot as a crutch when using it
    to write code and explain code. Overall, I liked working with Copilot, but
    my work habits \textbf{caused me to rely on it too much sometimes}.}{P 25, End}
\end{itemize}

The sentiments of Copilot being both helpful and hindersome to learning reflect
the tension that exists between helping students gain proficiency with LLM
usage and providing students the foundational skills needed to code
independently. 

\subsection{RQ3) {Demographic Outcomes}}\label{sec:perf_CS1LLM_demographics}

In this section, we evaluate how students in different demographic groups in our CS1-LLM
course compare in their performance outcomes.  Some gaps in outcomes between groups have
been studied extensively in the CS education community, as described in Section
\ref{subsec:student_outcomes}, and we analyze demographics here to explore how
the gaps in performance outcomes that may exist in our GenAI-centered course
differ from historical trends.

\subsubsection{Data Collection and Analysis}\label{sec:cs1llm_data_collection_analysis}

We focus on exam and project performance outcomes:
\begin{enumerate}
    \item \textit{\textbf{Average summative exam grade:}} The average of all four mid-quarter exams and the final exam.
    \item \textit{\textbf{Average project grade:}} The average of the three open-ended projects given to students throughout the semester.
    \end{enumerate}
See Figure~\ref{fig:exam_proj} for the distribution of scores.
 To analyze how these performance outcomes statistically differ, we performed
two Ordinary Least Squares (OLS) regressions: once for average exam
score and once for average project score.
\begin{align*}
  \text{Grade} = \beta_0 & + \beta_1 \hlNonMale{\text{Non-Male}}
                            + \beta_2 \hlFirstGen{\text{First Gen}}  \\
                           & + \beta_3 \hlLowIncome{\text{Low Income}}  
                            + \beta_4 \hlESL{\text{ESL}}  \\
                           & + \beta_5 \hlPriorExp{\text{Prior Experience}}  
                            + \beta_6 \hlCSCE{\text{Computing Major}}  \\
                           & + \beta_7 \hlBLNPI{\text{BLNPI}}  
\end{align*} 
Here, each of the outcome variables was coded as a binary variable for:
\hlNonMale{Non-Male} whether the student identified as non-male (female or non-binary), 
\hlFirstGen{First Generation} if they were a first generation college student, 
\hlLowIncome{Low Income} if they were a low-income student as determined by Pell Grant eligibility, 
\hlESL{ESL} if they learned English as a second language, 
\hlPriorExp{Prior Experience} if they had prior programming experience, 
\hlCSCE{Computing Major} if they were in a computing-focused major, and 
\hlBLNPI{BLNPI} if they identified as Black, Latine/x/a/o, Native American, or Pacific Islander. After
filtering out students who did not complete the final exam, project, or demographic
survey, we were left with a total of 207 students. 

\subsubsection{Grade Outcomes on Exams}

The results of the exam grades regression in Table~\ref{tab:grade_proj_ols} show that certain demographic
groups were more likely to have significantly higher or lower exam scores
including \hlBLNPI{BLNPI students} ($\beta_7=-8.40$, $p<0.001$) and students
with \hlPriorExp{prior experience} ($\beta_5=3.19$, $p=0.049$). 

\subsubsection{Grade Outcomes on Projects}
The results of the project grades regression in
Table~\ref{tab:grade_proj_ols} show that performance is not at all predicted by
demographics ($R^2_{adj}$ = -0.011)---likely because of a ceiling effect on
project grades. Given the plentiful high-level scaffolding and access to
low-level programming help from Copilot throughout the process of project completion, it may
be the case that there was sufficient infrastructure available for students of
all backgrounds to successfully create the projects.  

\section{Discussion}\label{sec:discussion}
Our study provides an initial examination of student outcomes and perceptions
in a CS1 course that fully incorporates LLMs. In summary, our core findings are:
\begin{enumerate}
  \item \textbf{Students performed similarly to previous offerings of a CS1}: Despite differing course goals and modified learning objectives, students in our course performed comparably to students in previous offerings of a CS1 course on fundamental programming skills such as code tracing and code explaining. This is encouraging for our community's general concern of students continuing to learn fundamentals of programming.
  \item \textbf{Students reported positive experiences with Copilot with some concerns}: While the vast majority of students reflected positively with respect to learning and productivity, some expressed concerns on the implications of overreliance.
 
  \item \textbf{Some existing equity gaps persist in exam scores}: Performance disparities by race and prior experience remained on exams (Table ~\ref{tab:scores}), reflecting historical findings. 
  \item \textbf{Projects mitigate equity gaps}: Project grades did not reflect the historical equity gaps found in exams (Table ~\ref{tab:scores}). We found {\it no} significant differences between historically underrepresented groups and highly represented groups on projects.
   \end{enumerate}
In this section, we explore implications of our findings for educators, curriculum designers, and the broader  research community.

\subsection{Implications for educators}
Our results suggest the need for \textit{clearer communication to students} on expected mastery of programming without access to Copilot both within the course \textit{and} in the world of professional programming. Helping students calibrate with respect to these expected outcomes is essential in ensuring that they perceive what they learn in the course as being reflective of what ``real programmers'' both know and do. This is important in establishing and bolstering students' self-assessments of themselves as ``real programmers'' and, as such, carrying with them a sense of belonging~\cite{krause2021relationship} and self efficacy~\cite{lishinski2016learning}. Future work should consider additional mechanisms by which these expectations can be communicated to students and possible alterations to the curriculum that can facilitate this.

Additionally, the interface between CS1 courses and follow-on courses will likely impact overall student success and which subgroups of students succeed.  If subsequent courses ignore that the learning outcomes have changed in CS1, students may unnecessarily struggle on some topics while being overprepared on others. For example, in our course, we de-emphasize writing code from scratch (because we believe that this skill now has a diminished importance); instructors within a department need to communicate to ensure smooth transitions between courses.

Our finding that projects mitigate historical equity gaps is a promising one, and suggests that educators may consider incorporating open-ended projects like ours that allow LLM usage. While exams remain a source of tension for many groups in introductory computing courses, projects may provide an assessment environment in which students from a variety of backgrounds can thrive. In fact, returning to our learning objectives in Section~\ref{subsec:llm_learning_objectives}, projects, rather than exams, may better target our higher-level objectives of problem decomposition, debugging, and writing complete programs.

\subsection{Implications for Course Designers and Researchers}

In the study by \citet{lau2023ban}, the authors found that instructors
maintained one of two perspectives toward teaching with GenAI. Some instructors
wanted to ban GenAI tools, feeling that using GenAI would lead to an erosion of
``fundamentals''. Other instructors wanted to embrace AI tools for reasons
similar to our own: to prepare students for future jobs and teach more advanced
material. Our study may offer useful insights into this ongoing ban vs. embrace
debate in favor of embracing GenAI---particularly with respect to the more
complex and open-ended assignments students were able to complete under this
approach.

One open question for the community that was exposed by our study is whether it
is important for students to continue to code from scratch in introductory
courses now that LLMs can write that code.
In our course, roughly half of the grade was for performance on non-GenAI exam questions.  
This reflects the belief that there is still something to be gained by writing code
from scratch, perhaps in terms of building student mental models of state and
execution. In fact, through our surveys (in particular, see Theme 4 in our
qualitative analysis), we discovered that students, too, want to be able to
code without help from GenAI. How much should students code from scratch?
Should they write more code early in the term and progressively off-load it to
the LLM? These are important questions for the broader research community.

\section{Limitations}

This study was conducted in one course at one institution.  The depth of
comparisons we can make between this course and the international average are
limited as this institution may under- or over-perform relative to that
average.  Only 400 students out of 556 completed the mid-quarter survey, 315
completed the end-of-quarter survey, and 207 completed the demographic
information. Lastly, as this is a new course, comparisons with existing
GenAI-based courses are limited.

\section{Conclusion}
This study examines a novel CS1 course integrating GenAI into
introductory programming, developed in response to the growing adoption of GenAI by
students, educators, and professionals. We assessed student performance
outcomes as well as student perceptions toward GenAI models. We find that
students performed similarly in CS1-LLM compared with earlier groups of
students in traditional CS1 courses on questions that involved code reading on
shared learning outcomes. The same groups who struggle in
traditional CS1 courses struggled on the exams in this course. By contrast, when students were permitted to use GenAI on large open-ended projects, we saw no such achievement gap. 
We note that these projects are larger and more complex than those generally assigned to traditional CS1 students, and that they closely match our higher-level course learning outcomes.
Student attitudes toward Copilot are mostly positive and indicate that they found the
tool helpful, although some expressed feelings of overreliance on it.

\vspace{-0.4em}




\bibliographystyle{ACM-Reference-Format}
\bibliography{sn-bibliography}

\end{document}